# Entanglement in Disordered Systems at Criticality

**Imre Varga**[*,1,2] **and José Antonio Méndez-Bermúdez**[3]

[1] Elméleti Fizika Tanszék, Fizikai Intézet, Budapesti Műszaki és Gazdaságtudományi Egyetem, Budafoki út 8. H-1111 Budapest, Hungary
[2] Fachbereich Physik, Philipps Universität Marburg, Renthof 5, D-35037, Marburg an der Lahn, Germany
[3] Instituto de Física, Universidad Autónoma de Puebla, Apartado Postal J-48, Puebla 72570, Mexico

[*] Corresponding author: e-mail varga@phy.bme.hu, Phone: +36 1 463 4109, Fax: +36 1 463 3567

Entanglement is a physical resource of a quantum system just like mass, charge or energy. Moreover it is an essential tool for many purposes of nowadays quantum information processing, e.g. quantum teleportation, quantum cryptography or quantum computation. In this work we investigate an extended system of $N$ qubits. In our system a qubit is the absence or presence of an electron at a site of a tight-binding system. Several measures of entanglement between a given qubit and the rest of the system and also the entanglement between two qubits and the rest of the system is calculated in a one-electron picture in the presence of disorder. We invoke the power law band random matrix model which even in one dimension is able to produce multifractal states that fluctuate at all length scales. The concurrence, the tangle and the entanglement entropy all show interesting scaling properties.

## 1 Introduction

The concept of entanglement [1] is believed to play an essential role in quantum information processing [2]. It lies in the heart of quantum mechanics, it is a manifestation of the genuinely quantum correlations. For instance quantum teleportation, a technique for moving a quantum state between two places requires the use of entangled states. The characterization of entanglement occurring in diverse systems both sheds new light on the detailed properties of the systems and explores the applicability of the systems in terms of quantum communication and quantum computing.

A number of characteristics have been developed in order to characterize entanglement in a bipartite or a multipartite system. In many cases systems with a small Hilbert space were considered. Due to the existence of an appropriate isomorphism between $N$ qubits (spins) and the many body systems of $N$ fermions extended systems start to be in the frontline of the investigation. Among them recently a number of authors studied the effect of quasiperiodicity [3] or disorder [4]. The spatial extent of the eigenstates plays a crucial role in the entangling properties of the system. Here we study the effect of critical, i.e. multifractal wave functions that occur at the metal-insulator transition in the Anderson problem [5].

## 2 Entanglement measures

Given a closed system described by a hamiltonian $H$, we may divide it into subsystems $A$ and $B$. A pure state is not separable, i.e. entangled, if it is not possible to decompose it simply over the direct product of $A$ and $B$. In other words the Schmidt decomposition of a state $|\Psi\rangle$ consists of more than one term [2]. Then the entanglement may be quantified using several measures.

The entanglement entropy is simply the von Neumann entropy of the reduced density matrix of subsystem $A$, $\rho_A = \text{Tr}_B\{|\Psi\rangle\langle\Psi|\}$,

$$E(\rho_A) = -\text{Tr}\{\rho_A \log(\rho_A)\} \qquad (1)$$

This quantity is identical to $E(\rho_B)$. Another quantity has been introduced by Meyer and Wallach [6] and has been used widely in the case of multipartite systems. It is called tangle and is directly related to the quantity that measures the mixedness or the purity of $\rho_A$. This quantity is defined as

$$Q(\rho_A) = 2(1 - \text{Tr}\rho_A^2) \qquad (2)$$

which in fact is another entropy, the linear entropy.

In the case of a bipartite system with two qubits, one of the most popular measures of entanglement is the concurrence introduced by Wootters in [7]. It is obtained from the eigenvalues of matrix

$$R_A = \rho_A(\sigma_y \otimes \sigma_y)\rho_A^*(\sigma_y \otimes \sigma_y).$$

The sorted eigenvalues $\lambda_1$, $\lambda_2$, $\lambda_3$ and $\lambda_4$ form the concurrence the following way

$$C(\rho_A) = \max(0, \sqrt{\lambda_1} - \sqrt{\lambda_2} - \sqrt{\lambda_3} - \sqrt{\lambda_4}). \quad (3)$$

All the above measures of entanglement vanish for separable states and attain their maximum for maximally entangled states.

### 3 Entanglement in extended systems

A linear chain consisting of $N$ sites serve as the system where entanglement is investigated. Each site may be occupied by an electron therefore the absence or presence of a particle on a site behaves as a qubit. The $N$ qubit system has a Hilbert-space of $2^N$ elements. In the present work we restrict ourselves to the single-particle problem reducing to an $N$-dimensional subset of the Hilbert-space. A given state will be expanded in this space as

$$|\Psi\rangle = \psi_1|100...0\rangle + \psi_2|010...0\rangle + ... + \psi_N|000...1\rangle \quad (4)$$

In the case of analyzing a single qubit, say the $i$th one, out of the whole system the natural basis is $|1\rangle_A = |0\rangle$ and $|2\rangle_A = c_i^+|0\rangle$, spanning a two dimensional space, where $|0\rangle_A$ is the vacuum state. In this case the reduced density matrix of a pure state $|\Psi\rangle$ is

$$\rho_i = \begin{pmatrix} 1-q_i & 0 \\ 0 & q_i \end{pmatrix}, \quad (5)$$

Where $q_i = \langle n_i \rangle = |\psi_i|^2$ is the occupation number at site $i$ in the given state. A straightforward analysis gives the reduced density matrix for a two-qubit subsystem yielding the so-called standard density matrix

$$\rho_i = \begin{pmatrix} v & 0 & 0 & 0 \\ 0 & w & z & 0 \\ 0 & \bar{z} & u & 0 \\ 0 & 0 & 0 & y \end{pmatrix}, \quad (6)$$

where the non-vanishing elements of the density matrix are given as $v = 1 - \langle n_i \rangle - \langle n_j \rangle + y$, $w = \langle n_i \rangle - y$, $u = \langle n_j \rangle - y$, $y = \langle n_i n_j \rangle$, $z = \langle c_i^+ c_j \rangle$. In our case we put one single electron into the lattice therefore we have $y = 0$.

The concurrence can be calculated for the above case of two qubits yielding the following simple form

$$C_{ij} = 2|\psi_i \psi_j|. \quad (7)$$

In the case the state $|\Psi\rangle$ is an eigenstate $|\mu\rangle$ of the system we may calculate average values of these entanglement measures. For instance the average concurrence can be readily calculated as

$$\langle C^\mu \rangle = \frac{1}{M}\sum_{i<j} C_{ij} = \frac{1}{M}\left[\left(\sum_i |\psi_i^\mu|\right)^2 - 1\right] \quad (8)$$

where $M = N(N-1)/2$. Moreover the average of the square of the concurrence can be averaged to give

$$\langle (C^\mu)^2 \rangle = \frac{2}{M}\left(1 - D_\mu^{-1}\right) \quad (9)$$

where

$$D_\mu = \left(\sum_i |\psi_i^\mu|^2\right)^{-1} \quad (10)$$

is the delocalization measure (participation number) of the eigenstate. For extreme localization $D_\mu = 1$, then the l.h.s of Eq. (9) vanishes because the state is separated onto one of the qubit-pairs, hence it is not entangled. For complete delocalization $D_\mu = N$, therefore the maximum value of entanglement is obtained.

An appropriate average value of (2) gives [4]

$$\langle Q^\mu \rangle = \frac{4}{N}\left(1 - D_\mu^{-1}\right). \quad (11)$$

Therefore both tangles, the average of the square of the concurrence and the Meyer-Wallach measure are simply related to the spatial localization properties of the states. As for the entanglement entropy we were unable to obtain a closed form as simple as Eqs. (10) and (11).

### 4 Disordered model systems at criticality

The system we consider here is a random matrix model in $d = 1$. The matrix elements are drawn from a random distribution with zero mean, $\langle H_{ij} \rangle = 0$, and a variance that decays in a power law fashion as a function of the distance from the diagonal

$$\langle |H_{ij}|^2 \rangle \propto \left(1 + \left(\frac{\sin(\pi|i-j|/N)}{\pi b/N}\right)^{2\alpha}\right)^{-1} \quad (12)$$

In the limit $1 \ll |i-j| \ll N$ this means that the variance behaves as

$$\langle |H_{ij}|^2 \rangle \propto \left(\frac{b}{|i-j|}\right)^{2\alpha}.$$

The spectral fluctuations and the properties of the wave functions share many characteristics with those of the Anderson model [5, 8]. The model for $\alpha \ll 1$ has a limit like a conventional random matrix ensemble with correlated spectrum and extended states, whereas for $\alpha \gg 1$ it behaves like a banded random matrix ensemble with bandwidth $b$, in which case as $N \to \infty$, the levels become uncorrelated and the corresponding eigenstates are localized. The value $\alpha = 1$ is marginal and then the system shows the behaviour just like that observed at the Anderson transition in $d \geq 3$: the spectra still show level repulsion but with also signs of no correlations and the eigenstates are multifractals. On the one hand this model is one dimensional, therefore linear scaling with system size is numerically easier. On the other hand, at $\alpha = 1$ we have not only a critical point but a continuous line of criticality parameterized with the effective bandwidth $b$. This parameter acts similarly as the critical conductance for the conventional Anderson transition. For $b >> 1.0$ we have again a limit similar to the one described by random matrix theory. This is the regime accessible to analytical calculations based on a nonlinear σ-model [8]. For $b << 1.0$, on the other hand, a new type of virial expansion has been developed in [9]. Very recently the supersymmetry method has also been applied in this regime [10].

In our numerical simulation we have diagonalized matrices of size $N = 128,\ldots,2048$. The average values of the entanglement measures have been further averaged over a window around the center of the band and over many realizations yielding high accuracy statistics.

### 4 Results and discussion
#### 4.1 Fixed *b*, varying α

As a first step we show the behaviour of the entanglement measures as a function of parameter $\alpha$ keeping parameter $b$ fixed. Since at $\alpha = 1$ a phase transition occurs we expect scale invariance, as well. As it is already noted in [3] for the states of a random matrix taken from the orthogonal class the average concurrence is

$$\langle C \rangle^{RMT} = \frac{4}{\pi} N^{-1} \tag{13}$$

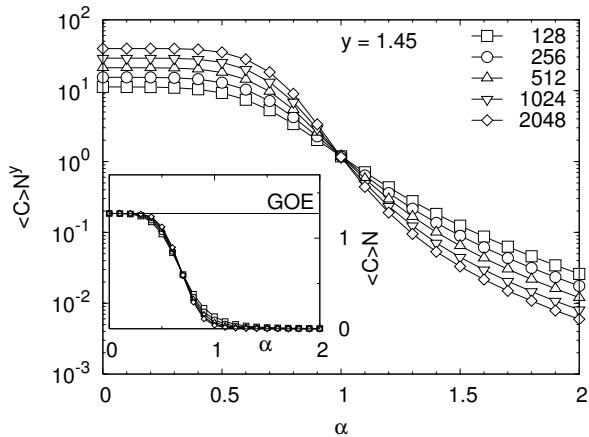

**Figure 1** For $b = 0.1$ scaling of the concurrence with $N$ as a function of α. The inset shows that as $\alpha \to 0$ the concurrence approaches its RMT value.

Therefore in Fig. 1 the inset shows that indeed as $\alpha \to 0$, we have $\langle C \rangle N = 4/\pi$. On the other hand the main panel shows that a scaling with $N$ with an exponent between 1 and 2 does provide scale invariance, as expected, at $\alpha = 1$.

The value of the exponent $y$ as a function of $b$ shows a fast decrease towards unity. A few more calculations allow us to deduce that as long as $b \geq 0.1$ the exponent $y$ seems to behave as

$$y - 1 \propto b^{-1}. \tag{14}$$

On the other hand as $b \to 0$, $y \to 2$ is expected [3].

The entanglement entropy shows similar behaviour as given in Fig. 2. The inset in Fig. 2 is the average single particle entropy of the states divided by $\log(N)$, which shows scale invariance at the transition point. Its value is the information dimension $D_1$.

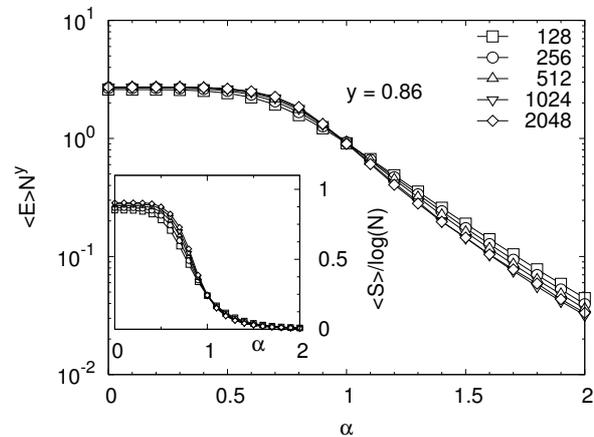

**Figure 2** For $b = 1.0$ scaling of the entropy of entanglement with $N$ as a function of α. The inset shows the scaling of the information dimension $D_1$.

#### 4.2 Behaviour at the critical point (α=1)

Next we investigate how the system behaves at criticality. We will present results of concurrence and tangle. A detailed analysis is left for a future publication.

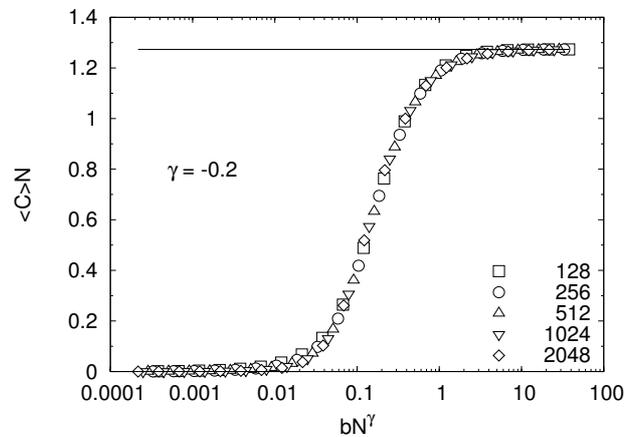

**Figure 3** Universal scaling of concurrence (3) at criticality.

Both the concurrence and the tangle are expected to be monotonous functions of $b$. In Figs. 3 and 4 we can clearly see a universal behaviour of the form

$$\langle X \rangle = N^{-1} f_X(N^\gamma b),  \quad (15)$$

Where $X$ stands for either $C$ or $Q$. The exponents with a yet unknown origin are different even in sign. The form of universal scaling, Eq. (15) resemble a recent finding of [11] in which fractal nature of the von Neumann entropy was found for quantum phase transitions, like the Anderson transition.

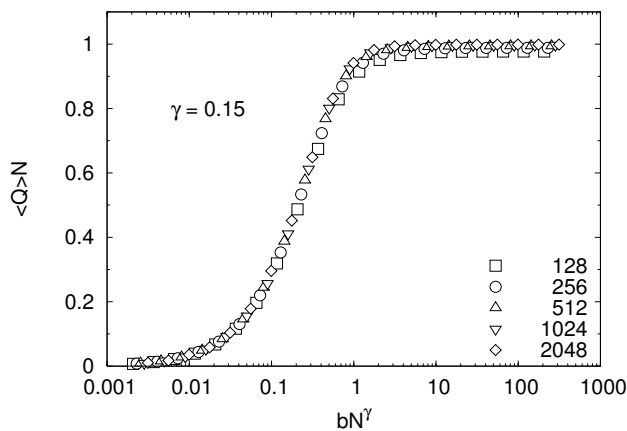

**Figure 4** Universal scaling of the tangle (2) at criticality.

The energy resolution of the entanglement measures may also give interesting universal scaling curves. In Fig. 5 we plotted the weighted density of states using the tangle quantities defined as

$$Q(E) = \left\langle \sum_\mu Q_\mu \delta(E - E_\mu) \right\rangle. \quad (16)$$

A scaling of the form

$$Q(E) = N^{-\nu} h(E\sqrt{N}) \quad (17)$$

was found with $\nu = 0.48$.

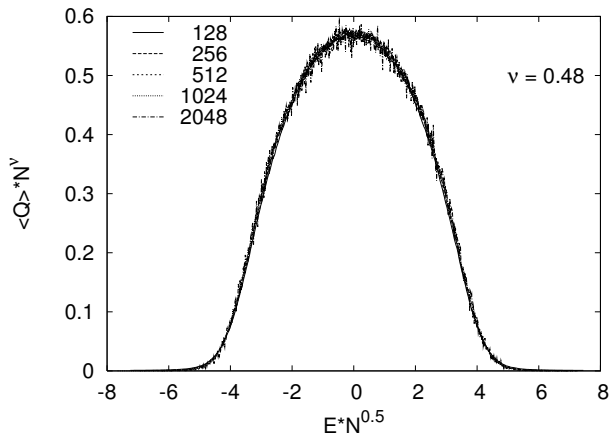

**Figure 5** Universal scaling of the energy resolution of the tangle according to Eq. (17). Here $b = 1.0$.

## 5 Conclusions

We have analyzed the entanglement properties of qubits arranged in a linear chain forming a ring where the intrinsic eigenstates change their nature according to a localization-delocalization transition characterized by multifractality. We have obtained universal scaling functions and exponents that describe the system right at the transition. For the understanding of these results substantial numerical experiments and analytical results are highly desirable.

Entanglement is a new physical resource that may be analyzed in order to understand systems and physical models. Beyond this advantage entanglement is obviously the tool to exploit the quantum correlations inherent in different systems that could be the basis for designing quantum computers. In this respect similar investigations are necessary to get a deeper insight into these issues.

**Acknowledgements** One of the authors (I.V.) is indebted to P. Lévay for fruitful discussions. Financial support from the Alexander von Humboldt Foundation and the Hungarian Research Fund OTKA under contract T46303 are gratefully acknowledged.

## References

[1] E. Schrödinger, Naturwissenschaften **23** 809.
[2] R. Horodecki, P. Horodecki, M. Horodecki, and K. Horodecki, arXiv:quant-ph/0702225; M.B. Plenio and S. Virmani, Quant. Inf. Comp. **7,** 1 (2007); L. Amico, R. Fazio, A. Osterloh, and V. Vedral, arXiv:quant-ph/07030441; W.K. Wootters, Quant. Info. And Comp. **1**, 27 (2001).
[3] A. Lakshminarayan and V. Subrahmanyam, Phys. Rev. A **67,** 052304 (2003).
[4] H. Li, X.G. Wang, and B. Hu, J. Phys. A: Math. Gen. **37,** 10665 (2004); H. Li and X.G. Wang, Mod. Phys. Lett. B **19,** 517 (2005); R. López-Sandoval and M.E. Garcia, Phys. Rev. B **74,** 174204 (2006).
[5] F. Evers and A.D. Mirlin, arXiv:0707.4378; C. DiCastro and R. Raimondi, in: Proceedings of the International School of Physics "Enrico Fermi", Varenna, Italy, 2003, (IOS Press, Bologna, 2004), pp. 259-333;
[6] D.A. Meyer and N.R. Wallach, J. Math. Phys. **43**, 4273 (2002).
[7] S. Hill and W.K. Wootters, Phys. Rev. Lett. **78,** 5022 (1997); W.K. Wootters, Phys. Rev. Lett. **80**, 2245 (1998).
[8] A.D. Mirlin, Y.V. Fyodorov, F.-M. Dittes, J. Quezada, and T.H. Seligman, Phys. Rev. B **54**, 3221 (1996); A.D. Mirlin, Phys. Rep. **326**, 259 (2000); E. Cuevas, M. Ortuno, V. Gasparian, and A. Perez-Garrido, Phys. Rev. Lett. B **88,** 016401 (2002); I. Varga, Phys. Rev. B **66**, 094201 (2002).
[9] O. Yevtushenko and V.E. Kravtsov, J. Phys. A: Math. Gen. **36**, 8265 (2003).
[10] O. Yevtushenko and A. Ossipov, J. Phys. A: Math. Gen. **40**, 4691 (2007).
[11] A. Kopp, X. Jia, and S. Chakravarty, Ann. Phys. **322**, 1466 (2007).